
\input harvmac
\noblackbox


\let\useblackboard=\iftrue
%
%
\newfam\black

\useblackboard
\message{If you do not have msbm (blackboard bold) fonts,}
\message{change the option at the top of the tex file.}
\font\blackboard=msbm10 scaled \magstep1
\font\blackboards=msbm7
\font\blackboardss=msbm5
\textfont\black=\blackboard
\scriptfont\black=\blackboards
\scriptscriptfont\black=\blackboardss

\else

\fi
%
\def\yboxit#1#2{\vbox{\hrule height #1 \hbox{\vrule width #1
\vbox{#2}\vrule width #1 }\hrule height #1 }}
\def\fillbox#1{\hbox to #1{\vbox to #1{\vfil}\hfil}}
\def\ybox{{\lower 1.3pt \yboxit{0.4pt}{\fillbox{8pt}}\hskip-0.2pt}}
%
%

\def\comments#1{}

\def\II{\relax{I\kern-.10em I}}

\def\IZ{\relax\ifmmode\mathchoice
{\hbox{\cmss Z\kern-.4em Z}}{\hbox{\cmss Z\kern-.4em Z}}
{\lower.9pt\hbox{\cmsss Z\kern-.4em Z}}
{\lower1.2pt\hbox{\cmsss Z\kern-.4em Z}}
\else{\cmss Z\kern-.4emZ}\fi}
\def\IB{\relax{\rm I\kern-.18em B}}
\def\IC{{\relax\hbox{$\inbar\kern-.3em{\rm C}$}}}
\def\ID{\relax{\rm I\kern-.18em D}}
\def\IE{\relax{\rm I\kern-.18em E}}
\def\IF{\relax{\rm I\kern-.18em F}}
\def\IG{\relax\hbox{$\inbar\kern-.3em{\rm G}$}}
\def\IGa{\relax\hbox{${\rm I}\kern-.18em\Gamma$}}
\def\IH{\relax{\rm I\kern-.18em H}}
\def\II{\relax{\rm I\kern-.18em I}}
\def\IK{\relax{\rm I\kern-.18em K}}
\def\IP{\relax{\rm I\kern-.18em P}}

%

\def\inbar{\,\vrule height1.5ex width.4pt depth0pt}

\font\cmss=cmss10 
\def\IR{\relax{\rm I\kern-.18em R}}

%


%

\def\lp10{\ell_p^{10}}
\def\lp11{\ell_p^{11}}
\def\R11{R_{11}}

\def\frac#1#2{{#1 \over #2}}

\def\mfl{(-1)^{F_L}}


\hyphenation{Di-men-sion-al}



%
\lref\JoyceI{D.D. Joyce, ``Compact Riemannian 7-manifolds with
$G_2$ Holonomy, I," J. Diff. Geom. {\bf 43} (1996) 291.}

\lref\JoyceII{D.D. Joyce, ``Compact Riemannian 7-manifolds with
$G_2$ Holonomy, II," J. Diff. Geom. {\bf 43} (1996) 329.}

\lref\SenF{A. Sen, ``F-theory and Orientifolds," Nucl. Phys.
{\bf B475} (1996) 562;
A. Sen, ``Orientifold Limit of F-theory Vacua,'' Phys. Rev.
{\bf D55} (1997) 7345, hep-th/9702165.}

\lref\SenEn{A. Sen, ``A Note on Enhanced Gauge Symmetries in
M and String Theory,'' JHEP {\bf 9709} (1997) 001, hep-th/9707123.}

\lref\Gtwocon{H. Partouche and B. Pioline, ``Rolling Among
$G_2$ Vacua,'' JHEP {\bf 0103} (2001) 005, hep-th/0011130.}

\lref\AMV{M. Atiyah, J. Maldacena and C. Vafa, ``An M-theory Flop
as a Large N Duality,'' hep-th/0011256.}

\lref\SinVa{S. Sinha and C. Vafa, ``SO and Sp Chern-Simons at Large N,''
hep-th/0012136.}

\lref\Cachazo{F. Cachazo, K. Intriligator and C. Vafa,
``A Large N Duality via a Geometric Transition,''  hep-th/0103067.}

\lref\Acharya{B.S. Acharya, ``On Realizing ${\cal N}=1$ Super Yang-Mills
in M-theory,'' hep-th/0011089;
``Confining Strings from G(2) Holonomy Space-Times,''
hep-th/0101206\semi
B.S. Acharya and C. Vafa, ``On Domain Walls of ${\cal N}=1$ Supersymmetric
Yang-Mills in Four-Dimensions,'' hep-th/0103011.}

\lref\Achold{B.S. Acharya, ``M-theory, Joyce Orbifolds and
Super Yang-Mills,'' Adv. Th. Math. Phys. {\bf 3} (1999) 227, hep-th/9812205.}

\lref\KLM{A. Klemm, W. Lerche and P. Mayr, ``K3 Fibrations and
Heterotic-Type II String Duality,'' Phys. Lett. {\bf B357} (1995) 313,
hep-th/9506112.}

\lref\Cvetic{ M. Cvetic, G.W. Gibbons, H. Lu, C.N. Pope,
``Supersymmetric Non-singular Fractional D2-branes and NS-NS 2-branes,''
hep-th/0101096;
``Hyper-Kahler Calabi Metrics, $L^2$ Harmonic Forms, Resolved M2-branes,
and
$AdS_4/CFT_3$ Correspondence,'' hep-th/0102185.}

\lref\Gomis{J. Gomis, ``D-branes, Holonomy and M-theory,''
hep-th/0103115.}

\lref\douglas{M. Douglas, ``Branes within branes,'' hep-th/9512077.}

\lref\chir{S. Kachru and E. Silverstein, ``Chirality Changing Phase
Transitions in 4d String Vacua,'' Nucl. Phys. {\bf B504} (1997) 272,
hep-th/9704185\semi
B. Ovrut, T. Pantev and J. Park, ``Small Instanton Transitions in
Heterotic M-theory,'' hep-th/0001133.}

\lref\SV{S. Shatashvili and C. Vafa, ``Superstrings and Manifolds
of Exceptional Holonomy,'' hep-th/9407025.}

\lref\ov{H. Ooguri and C. Vafa, ``Knot Invariants and
Topological Strings,'' Nucl. Phys. {\bf B577} (2000) 419,
hep-th/9912123.}

\lref\av{M. Aganagic and C. Vafa, ``Mirror Symmetry, D-branes
and Counting Holomorphic Discs,'' hep-th/0012041.}

\lref\GM{R. Gopakumar and S. Mukhi, ``Orbifold and Orientifold
Compactifications of F-theory and M-theory to Six and Four
Dimensions,'' Nucl. Phys. {\bf B479} (1996) 260, hep-th/9607057.}

\lref\HM{J. Harvey and G. Moore, ``Superpotentials and Membrane
Instantons,'' hep-th/9907026.}

\lref\vafawit{C. Vafa and E. Witten, ``Dual String Pairs with ${\cal N}=1$
and ${\cal N}=2$ Supersymmetry in Four Dimensions,'' Nucl. Phys.
Proc. Suppl. {\bf 46} (1996) 225, hep-th/9507050.}

\lref\KV{S. Kachru and C. Vafa, ``Exact Results for ${\cal N}=2$
Compactifications of Heterotic Strings,'' Nucl. Phys. {\bf B450}
(1995) 69, hep-th/9505105.}

\lref\ednun{J. Edelstein and C. Nunez, ``D6 branes and M theory
geometrical transitions from gauged supergravity,'' hep-th/0103167.}

\lref\Grossfest{E. Witten, ``M-theory Dynamics on Manifolds of
$G_2$ Holonomy,'' talk presented March 3, 2001 at ``Heterotic Dreams and
Asymptotic Visions,'' the 60th birthday celebration for David Gross,
ITP, Santa Barbara.}

\lref\Paul{P. Aspinwall and M. Gross, ``Heterotic-Heterotic String
Duality and Multiple $K3$ Fibrations,'' Phys. Lett. {\bf B382} (1996)
81, hep-th/9602118.}

\lref\MorrVaf{D. Morrison and C. Vafa, ``Compactifications of F-theory
on Calabi-Yau Threefolds, I,'' Nucl. Phys. {\bf B473} (1996) 74,
hep-th/9602114.}

\lref\SYZ{A. Strominger, S.T. Yau and E. Zaslow, ``Mirror Symmetry
Is T-Duality,''
Nucl. Phys. {\bf B479}(1996) 243, hep-th/9606040.}

\lref\kklmII{S. Kachru, S. Katz, A. Lawrence and J. McGreevy,
``Mirror symmetry for open strings,''
Phys.Rev. {\bf D62} (2000) 126005, hep-th/0006047.}

\lref\kklmI{S. Kachru, S. Katz, A. Lawrence and J. McGreevy,
``Open string instantons and superpotentials,''
Phys.Rev. {\bf D62} (2000) 026001, hep-th/9912151.}

\Title{\vbox{\baselineskip12pt\hbox{hep-th/0103223}
\hbox{SLAC-PUB-8799}\hbox{SU-ITP-01/14}}}
{\vbox{
\centerline{M-theory on Manifolds of $G_2$ Holonomy}
\bigskip
\centerline{and Type IIA Orientifolds}
}}
\centerline{Shamit Kachru$^{1,2}$ and John McGreevy$^{1}$}
\bigskip
\bigskip
\centerline{$^{1}$ Department of Physics and SLAC}
\smallskip
\centerline{Stanford University}
\smallskip
\centerline{Stanford, CA 94305/94309}
\medskip
\centerline{$^{2}$ Institute for Theoretical Physics}
\smallskip
\centerline{University of California}
\smallskip
\centerline{Santa Barbara, CA 93106}
\bigskip
\bigskip
\noindent

We demonstrate
that M-theory compactifications on 7-manifolds of
$G_2$ holonomy, which yield 4d ${\cal N}=1$ supersymmetric systems,
often admit at special loci in
their moduli space
a description as type IIA orientifolds.
In this way, we are able to find new dualities
of special IIA orientifolds, including dualities which relate
orientifolds of IIA
strings on manifolds of different topology with different numbers
of wrapped D-branes.
We also discuss models which incorporate, in a natural way,
compact embeddings
of gauge theory/gravity dualities similar to those studied
in the recent work of Atiyah, Maldacena and Vafa.

\bigskip
\Date{March 2001}

\newsec{Introduction}

Compactifications of M-theory and string theory down to 4d ${\cal N}=1$
supersymmetry are of obvious interest.
The reduced supersymmetry is probably necessary for any contact with
real world physics.  It also allows for
richer phenomena than extended supersymmetry
and so provides a nice playground for theorists.

Generic methods of constructing such models include compactifying the
heterotic string on Calabi-Yau threefolds, F-theory on Calabi-Yau
fourfolds, and M-theory on 7-manifolds of $G_2$ holonomy.
Although many basic facts about all of these classes of compactifications
remain mysterious, perhaps the least is known about the last class,
since at least the others are amenable to attack using techniques of
complex geometry.

A large class of compact 7-manifolds with $G_2$ holonomy was constructed
by Joyce \refs{\JoyceI,\JoyceII}.  In this note, we make the simple
observation that M-theory compactified on many of these spaces admits,
at special loci in its moduli space, a description as an orientifold
of type IIA string theory compactified on a Calabi-Yau threefold.\foot{
We ignore the possibility of a membrane instanton generated superpotential
\HM\ in most of our discussion.  If such a potential exists, it would
provide a potential barrier between the large-volume M-theory and
perturbative type IIA limits. Our successful comparison of these limits
in many cases suggests that either such a potential is absent,
or the interpolation over the potential barrier is nevertheless physically
meaningful.}
This is reminiscent of the fact that F-theory models can be
reformulated, at special loci in their moduli space, as type IIB
orientifolds
\refs{\SenF,\GM}.

There are several different reasons this observation can be useful.  On
the
one hand, the orientifolds we discuss have a rather simple, solvable
structure, and so provide a very concrete handle on these models at
some special points in their moduli space.  On the other hand, as
we will show, a given $G_2$ space can admit $\it different$ type IIA
orientifold limits.  Thus, by studying limit points in the moduli
space of $G_2$ compactifications, we learn about non-perturbative
dualities
of IIA compactifications with ${\cal N}=1$ supersymmetry.  In particular,
we exhibit an example where orientifolds of type IIA on Calabi-Yau
spaces of different topology (and with different numbers of D-branes
and orientifold planes) are dual to each other.  On yet a third hand,
our construction ``globalizes'' interesting
gauge theory/gravity dualities similar to those encapsulated in the
local models of \AMV\ and \SinVa.

In \S2, we introduce the $G_2$ manifold $X$ which will be our
focus in part of this note.
In \S3, we show that various limit points in the moduli space of
M-theory on $X$
are well described by IIA orientifolds.  This observation allows us to
find non-perturbative duality symmetries of these orientifolds.
In \S4 we make some remarks about the extent to which our
analysis generalizes to other $G_2$ spaces, and
also provide a simple proof that a large class of IIA
orientifolds should have an ``M-theory lift'' to $G_2$ compactifications.
In \S5 we explain how gauge theory/gravity
dualities analogous to those discussed in \refs{\AMV,\SinVa}
naturally arise
in simple examples of compact $G_2$ manifolds and the
related IIA orientifolds.
We conclude in \S6 by mentioning some interesting directions for further
study.

Several papers analyzing various related aspects of M-theory on
$G_2$ manifolds have appeared recently.
Dual descriptions of ${\cal N}=1$ gauge theories using such
spaces have been discussed
in \refs{\AMV, \SinVa, \Acharya, \Cachazo,\Grossfest,\ednun}, while
\Gomis\ discusses a general relationship
between certain classes of wrapped branes and geometries with exceptional
holonomy. Earlier work on this subject appears in \Achold.
Phase transitions between topologically distinct $G_2$
compactifications were described in \Gtwocon.

\newsec{The Manifold $X$}

A basic example of a compact 7-manifold of $G_2$ holonomy is
the manifold $X$ considered by
Joyce in \JoyceI.  It is constructed as a toroidal orbifold.  Let
$x_{1},\cdots,x_{7}$ parametrize a square $T^7$ which is a product
of seven circles of radii $r_1, \cdots, r_7$.
Define $X$ as the (desingularization
of the) quotient of this
$T^7$ by the $Z_2^3$ group with generators

\eqn\first{\alpha(x_i) = (-x_1, -x_2, -x_3, -x_4, x_5, x_6, x_7)}
\eqn\second{\beta(x_i) = (-x_1, 1/2 - x_2, x_3, x_4, -x_5, -x_6, x_7)}
\eqn\third{\gamma(x_i) = (1/2 - x_1, x_2, 1/2 - x_3, x_4, -x_5, x_6,
-x_7)}
where $1/2$ denotes a shift of order 2 around the circle.
Then as demonstrated in \JoyceI, $X$ has betti numbers $b_2(X) = 12,
~b_3(X) = 43$.
Therefore,
M-theory compactification on $X$ gives rise to a 4d ${\cal N}=1$
supersymmetric
low-energy theory with (generically) 12 abelian
vector multiplets and 43 chiral multiplet moduli.

It may be useful to review the origin of the various cohomology classes on
$X$ here.  None of the two-forms and seven of the
three-forms on $T^7$ are invariant
under the action of $\langle \alpha,\beta,\gamma \rangle$.\foot{$\langle
\cdots
\rangle$ denotes ``the group generated by $\cdots$.''}
In addition, each of
the generators fixes 16 $T^3$s on $T^7$; however e.g. the 16 $T^3$s fixed
by $\alpha$ are identified by the group $\langle \beta,\gamma \rangle$ to
yield
4 on the quotient $X$, and the fixed tori of $\beta$ and $\gamma$ undergo
a similar fate.
The local form of the singularities at the fixed $T^3$s is $R^4 / Z_2
\times
T^3$, and resolving each of these yields a two-form and three three-forms.
Since there are 12 such fixed tori on $X$, after desingularizing one has
the stated betti numbers.

\newsec{Orientifold Limits of $X$}

In this section, we demonstrate that $X$ has several different IIA
orientifold
limits in its moduli space.  This in particular tells
us that the different orientifolds
are related to one another by various dualities.

\subsec{Orientifold A}

We start by viewing $x_7$ as the ``M-theory circle,'' or the eleventh
dimension.  Then in the limit of small $r_7$, we should be able to get
an effective IIA description of M-theory on $X$.
Denote by $\alpha^*$ the action of $\alpha$ restricted to the
$T^6$ with coordinates $x_1, \cdots, x_6$.
Then since $\alpha$ and $\beta$ don't act on the M-theory circle anyway,
in the limit of small $x_7$ they simply induce identifications on the
$T^6$ visible to the type IIA string.  It is therefore propagating
on the Calabi-Yau space $N = T^6 / \langle \alpha^*,\beta^* \rangle$.

However, $\gamma$ also acts on the M-theory circle.  Using the results of
\SenEn, it follows that the action of $\gamma$ (inversion of the M-theory
circle and three other coordinates) is mapped in the IIA theory to
$w = (-1)^{F_L} ~\Omega~\gamma^*$.
Thus, IIA string theory on the orientifold of $N$ by $w$ should
govern M-theory in the limit of small radius for $x_7$.
Let us call this model orientifold A.

To check this conjecture, let us try to match up the counting of fields.
$N$ has hodge numbers $h^{1,1} = 19, h^{2,1} = 19$.
So IIA string theory on $N$ yields an
${\cal N}=2$ supersymmetric
4d theory with 20 hypermultiplets (including the dilaton)
and 19 vector multiplets.
Now, projecting by $w$ has the following effect (see e.g. \vafawit).
Each of the 20 hypermultiplets is projected down to a chiral multiplet.
The vector multiplets (which came from the K\"ahler moduli) are more
subtle: those which come from (1,1) forms $\it invariant$ under
$w$ give rise to ${\cal N}=1$ vector multiplets, while those which
are $\it anti-invariant$ under $w$ give rise to chiral multiplets.
It is easy to convince oneself that the untwisted (1,1) forms on $N$
are anti-invariant, while the 16 twisted (1,1) forms split into
$\pm$ eigenspaces of equal size.  Therefore, the K\"ahler moduli
contribute 11 chiral multiplets and 8 vector multiplets.

This accounts for 31 chiral multiplets and 8 abelian
vectors so far.  However,
we must also take into account the fixed points of the $w$ action.
$\gamma^*$ acts with 8 fixed loci on $T^6$. Identification
by $\langle \beta,\gamma \rangle$ reduces this to 2 fixed loci;
a neighborhood of each
in the threefold is of the form $R^3 / Z_2 \times T^3$.
Therefore, there are orientifold
six-planes wrapping each of these $T^3$s (as in \SenEn).

By the normal tadpole cancellation considerations, we must introduce
2 $D6$ branes for each $O6$ plane.  Hence, we introduce a total of
4 $D6$ branes wrapping $T^3$s in this model.
Each of the $D6$ branes comes with a $U(1)$ vector multiplet and
3 chiral multiplet moduli (coming from the Wilson lines on the
$T^3$, together with the moduli of the three-cycle in $N$).
So the $D6$ branes contribute a total of 12 chiral multiplet moduli
and 4 vector multiplets to the low-energy theory.

Totalling up the spectrum, we find that orientifold A (at generic
points in its moduli space) has 43 chiral
multiplet moduli and 12 vector multiplets, just as it must to match
the spectrum of M-theory on $X$.  At special points in
moduli space when the $D6$ branes coincide, one achieves
enhanced gauge symmetries, which come from geometrical singularities
in the M-theory picture \SenEn.

\subsec{Orientifold B}

It is of course also possible to view other circles as the M-theory
circle.  For instance, we could take $x_4$ to be the M-theory circle.
However, repeating the same logic as in \S3.1, we would find that
we again arrive (in the small $r_4$ limit) at an orientifold (which
we could call orientifold B) of
type IIA string theory on the Calabi-Yau orbifold with hodge numbers
$h^{1,1} = 19, h^{2,1} = 19$, and we again have to introduce the
same numbers of $D6$ branes.

The role of the dilaton in orientifold $A$ is played by a geometrical
modulus in orientifold $B$, and vice-versa.  However, they are really
compactifications on the same target space.
This means that the IIA theory on the orientifold of $N$ by $w$,
discussed in \S3.1, has
a sort of $S-T$ exchange symmetry, where $S$ is the dilaton
chiral multiplet and $T$ is the chiral multiplet containing the radius
of $x_4$.
This way of seeing the $S-T$ exchange symmetry of these orientifold
models is analogous to the way that the $S-T$ exchange symmetry \KLM\ of
the main heterotic string examples in \KV\
can be understood as
arising from the existence of multiple $K3$ fibrations in the
type II-dual Calabi-Yau compactifications \refs{\Paul,\MorrVaf}.

In fact, this model enjoys more symmetry than just a single $S-T$
duality; one could equivalently consider the $x_6$ circle to be
the M-theory circle, with the same results, yielding a sort of
$S-T-U$ triality symmetry.

\subsec{Orientifold C}

A more interesting possibility is to interpret $x_5$ as
the M-theory circle.
Then acting on the $T^6$
coordinates $x_1, x_2, x_3, x_4, x_6, x_7$, we have

\eqn\newalp{\alpha(x_i) = (-x_1, -x_2, -x_3, -x_4, x_6, x_7)}

\eqn\newbg{
\beta\gamma(x_i) = (x_1 + 1/2, 1/2 - x_2, -x_3, x_4, -x_6, -x_7)}
The manifold $N^\prime = T^6 / \langle \alpha, \beta\gamma \rangle$
is then a Calabi-Yau threefold with hodge numbers $h^{1,1} = 11, h^{2,1} =
11$.
In particular, it is topologically distinct from the threefold
$N$ which appeared in \S3.1\ and \S3.2.

Define $u$ to be the composition
of $(-1)^{F_L}\Omega$ with the action of $\gamma$ on the $T^6$
coordinates.
Then in the limit of small $r_5$, M-theory on $X$ should be well described
by IIA theory on
the orientifold of $N^\prime$ by $u$, which we will call orientifold C.

Let's check that the spectrum matches our expectations.
$v$, the composition of $(-1)^{F_L} \Omega$ with the
action of $\beta$ on the $T^6$ coordinates, arises upon
composing $u$ with elements of the orbifold group.
Both $u$ and $v$ act with fixed loci on $N^\prime$.  Each has
8 fixed $T^3$s in the $T^6$, which descend to 2 fixed $T^3$s
in the orbifold $N^\prime$.  Therefore, one has to introduce
four $O6$ planes, and 8 wrapped
$D6$ branes are required to cancel the
RR tadpoles.  These give rise to 8 abelian vector multiplets and
24 chiral multiplets, at generic points in moduli space.

The projection of the spectrum of IIA on $N^\prime$ can be done as
before.
Once again, half of the 8 twisted (1,1) forms are
invariant under the orientifold action, while the other
half (and the untwisted (1,1) forms) are anti-invariant.
So we get 4 vectors and 7 chirals from the (1,1) forms; adding
in the 12 chirals descending from the ${\cal N}=2$ hypermultiplets,
we indeed find a total of 43 chiral multiplets and 12 vectors.

In this orientifold C picture, the radii $r_4, r_6, r_7$ which are
related (up to triality) to the dilaton in the pictures of \S3.1\
and \S3.2\ are all geometrical moduli of the IIA compactification,
while $r_5$ (which is geometrized in orientifolds A,B) is playing
the role of the dilaton.
This gives an example of a strong/weak duality between IIA orientifolds of
topologically distinct Calabi-Yau spaces, with different numbers of
space-filling D-branes and orientifold planes.

\newsec{Generalization to Other Models}

In this section, we generalize our results in two directions.
We first show that a large class of $G_2$ spaces should
similarly have orientifold limits.  We then
take the opposite approach, and prove that a wide class of
IIA orientifolds have an M-theory lift to $G_2$ compactifications.

\subsec{Other Classes of $G_2$ Manifolds}

Beyond the toroidal orbifold constructions of Joyce, there are
other methods of constructing $G_2$ holonomy spaces which
are amenable to an orientifold interpretation.

\bigskip
\noindent{\it{Barely $G_2$ Manifolds}}

Harvey and Moore defined ``barely $G_2$ manifolds'' as quotients of the
form $X = (Y \times S^1) / Z_2$, where $Y$ is a Calabi-Yau threefold
and the $Z_2$ action is a composition of a freely
acting antiholomorphic involution $\sigma$ on $Y$ with inversion on
the circle factor $x_7$ \HM.  These are of course a special case of a more
general construction which should arise
when $\sigma$ has fixed points \refs{\JoyceI,
\JoyceII}.

For the barely $G_2$ spaces, it turns out that
\eqn\hthree{H^{3}(X) = H^{2}(Y)^{-} +  H^{3}(Y)^+}
\eqn\htwo{H^{2}(X) = H^{2}(Y)^{+}}
where $\pm$ refer to eigenvalues under the action of
$\sigma$ on $Y$.
For simple examples which come from hypersurfaces in toric
varieties, one simply keeps the complex structure deformations
which preserve the real structure (i.e. defining equations with
real coefficients), so
$H^3(Y)$ has $\pm$ eigenspaces of
equal dimension.  For such examples, we find
$n_C = h^{1,1}(Y)^- + h^{2,1}(Y) + 1$
chiral multiplets and $n_V = h^{1,1}(Y)^{+}$ vector multiplets
in M-theory on $X$.

As one shrinks the radius $r_7$ of the $S^1$, one should obtain a
IIA description.  Indeed, since the $Z_2$ above acts with an inversion
on $x_7$, we should expect that the orientifold of IIA on $Y$ by
$(-1)^{F_L} \Omega$ composed with $\sigma$
arises in this limit.
It follows from the general considerations of \vafawit\ (as discussed
in \S3) that the
spectrum of this type IIA orientifold agrees with the M-theory
spectrum.

\bigskip
\noindent{\it{Cases with Fixed Points}}

It is
attractive to speculate about generalizations of the previous case
to cases where $\sigma$ acts on $Y$ with fixed points.
On general grounds, the fixed point locus $\Sigma \subset Y$ will
be a special Lagrangian (sL) three-cycle (or several, in which case one
should
repeat the discussion below for each component).  It is not known in
generality
how to resolve the singularities in this case to obtain a smooth
metric of $G_2$ holonomy.

However, the existence of an orientifold limit leads to a very natural
conjecture.  Shrinking the $x_7$ circle again, we find a IIA model
which should have an $O6$ plane and two $D6$ branes wrapping $\Sigma$.
For $\Sigma$ special Lagrangian, a D6 brane wrapping $\Sigma$ gives
rise to a single ${\cal N}=1$ vector multiplet and $b_1(\Sigma)$
chiral multiplets in spacetime.  Therefore, we expect that there
will be 2 vectors and $2b_1(\Sigma)$ chiral multiplets associated
with the $D6$ branes in this limit.
When the $D6$ branes are coincident, the model has enhanced gauge
symmetry (which shows up in the M-theory as the singularity of
the $G_2$ space related to the fixed points of $\sigma$).  For
$b_1(\Sigma) > 0$, one can move in the $D6$ brane moduli space to
remove the enhanced gauge symmetry.  It is then attractive to conjecture
that in the M-theory picture, $b_1(\Sigma) > 0$ is a condition that
allows the singularities of this class of $G_2$ orbifolds to be repaired,
and that furthermore resolving the singularity gives rise to
precisely two elements of $b_2(X)$ and $2 b_1(\Sigma)$ elements
of $b_3(X)$.\foot{This could be related to Condition 4.3.1
in \JoyceII, which was stated without proof to be an important condition
in
resolving singularities of this sort.}

\subsec{``All'' Orientifolds of type IIA on CY have a $G_2$ Limit}

Suppose we have a IIA orientifold which gives rise to a four dimensional
${\cal N}=1$
supersymmetric theory.  For simplicity, let us first restrict ourselves to
orientifolds of tori.
The orbifold part of the
orientifold group
must have (at most) $SU(3)$ holonomy, to preserve (at least)
4d ${\cal N}=2$
supersymmetry.\foot{This is because there
are no geometric compactifications of IIA down to 4d which preserve
precisely 4d ${\cal N}=1$ supersymmetry.}
Let us assume we are in the most generic case, so
that it preserves precisely ${\cal N}=2$
supersymmetry.
Denote the full orientifold group by
\eqn\orgr{{\cal G} ~=~ \Gamma_1 \times (-1)^{F_L} \Omega \Gamma_2}
The $(-1)^{F_L}$ is present because we choose, as in \SenEn,
a convention
where reflection on three circles must be accompanied by a $(-1)^{F_L}$
to preserve supersymmetry in the IIA theory, and we will show momentarily
that all elements of $\Gamma_2$ must reflect precisely
three circles of the $T^6$.
With these assumptions, $T^6/\Gamma_1$ alone is Calabi-Yau, and so has a
holomorphic three-form $\Omega^{(3,0)}$ and a K\"ahler form $J$.

Now, consider the $\mfl \Omega \Gamma_2$ part of the group.
Any element
$\mfl \Omega g_2$ of this part must have $g_2$ reversing the orientation
of
the 6d target, or it cannot be a symmetry of the IIA theory.  So
we know a few things about the $g_2$ action:

\item{i)} $g_2$ maps $J$ to $-J$ (orientation reversal) and

\item{ii)} $g_2$ maps $\Omega^{(3,0)}$ to $\overline \Omega^{(0,3)}$.
Notice that this implies that $g_2$ reflects precisely three
circles of the $T^6/\Gamma_1$, as required above.

In $ii)$, we are using the fact that to preserve one
supersymmetry, some linear combination of the killing
spinors must be preserved.  This means that $g_2$ either preserves
the holomorphic and anti-holomorphic three-form
individually,
or at least preserves a
linear combination.
But since $\Omega^{(3,0)} \wedge \overline \Omega^{(0,3)} \sim J \wedge J
\wedge J$, by $i)$ above $g_2$ must permute the two.
One might worry that $g_2$ could act with a phase in
relating $\Omega^{(3,0)}$ to its conjugate; but
all $g_2 \subset \Gamma_2$ which exchange $\Omega^{(3,0)}$ and its
conjugate would have
to have
the same phase to preserve ${\cal N}=1$ supersymmetry.  It
can then be redefined to $1$
by a phase rotation of $\Omega^{(3,0)}$.

To proceed, we add a
seventh M-theory circle $x_7$.
Define the new group $\tilde {\cal G}$, which acts on $T^7$, as follows:
take each element of ${\cal G}$ and replace $\mfl \Omega$ anywhere it
appears with inversion of the $x_7$ coordinate (while elements which
don't include a $\mfl \Omega$ act trivially on the $x_7$ coordinate).
This will not change
anything about elements of $\Gamma_1$
(since the minus sign on $x_7$ will cancel in the
product of two $\mfl \Omega \Gamma_2$ elements).  However, under
assumptions i)
and ii) above, the three-form
\eqn\gtwostruct{ \Phi = J \wedge dx_7 + Re  [\Omega^{(3,0)}] }
is preserved by the whole (now orbifold) group $\tilde {\cal G}$
acting on $T^7$.
This form is preserved by a $G_2$ subgroup of
$GL(7,R)$ \JoyceII.  This is sufficient to
prove that the resulting manifold is a $G_2$ space.

It is clear that this argument is more generically applicable to
supersymmetric models which are not toroidal orientifolds.
One could replace the $T^6$ in the IIA theory with any manifold
$M$, use the fact that $M / \Gamma_1$ should be Calabi-Yau to
preserve supersymmetry in the IIA theory, and
apply the same logic.

\newsec{The Case of the Disappearing Orientifold}

Recent work has made it clear that gauge dynamics on wrapped
$D6$ branes (or arising from singular M-theory geometries) can
often be encoded by smooth geometries in a ``dual'' gravity
description \AMV.
The gauge dynamics is then encoded in appropriate RR-fluxes,
or in changes of the behavior of the M-theory three-form $C$
field, which (suitably interpreted)
capture the low-energy physics of the gauge theory.
In this section, we discuss examples of this phenomenon which
arise in string/M-theory compactifications in a natural way.

The most obvious source of
consistent compact models with wrapped $D6$ branes is the
Calabi-Yau orientifolds discussed here.
The components of the orientifold
fixed locus provide sL three-cycles
$\Sigma$, which are wrapped by orientifold planes and $D6$ branes.
In fact, examples of sL cycles $\Sigma$
which arise in this way were studied in \refs{\kklmI,\kklmII}
precisely
with the motivation of understanding the dynamics on the worldvolumes
of such wrapped $D6$ branes.

One interesting fact (which had perplexed some of the authors of
\refs{\kklmI,\kklmII} for some time)
is that it is possible for the fixed locus of an anti-holomorphic
involution to
disappear as the complex structure of the Calabi-Yau varies; and
the relevant complex structure moduli survive in the orientifold
models.
This fact was used in \kklmII\ to identify
D6 branes on such
real slices as mirror to D5 branes on
vanishing holomorphic curves.
However, it raises the question: if one continues past the
point in moduli space where the fixed locus disappears (so
there is no orientifold plane, and no need to introduce $D6$
branes), where
has the information about the gauge theory on the $D6$ branes
gone?  The gauge theory/gravity dualities relevant to this
situation were studied in \refs{\AMV,\SinVa}, and provide the answer to
this question.

Let us illustrate this with a simple example.  The easiest examples
discussed in \kklmII\ basically involve a
sL three-cycle which is the fixed locus of a real
involution and which collapses at a conifold singularity.  So locally,
the geometry of the compact Calabi-Yau $M$ looks like
\eqn\localsit{z_1^2 + z_2^2 + z_3^2 + z_4^2 ~=~\mu}
where $\mu$ is chosen to be a $\it positive$ real parameter.
Then under the involution
\eqn\invol{{\cal I}: z_i \to \overline z_i}
the fixed point locus
$\Sigma^{+}$ is the three-sphere
\eqn\sphere{\Sigma^{+}:~\sum_{i=1}^{4} x_i^2 ~= ~\mu}
where $z_i = x_i + i y_i$.

We can embed this situation in a $G_2$ manifold as in \S4.1, where the
$G_2$ manifold $X$ is of the form $(M \times S^1)/\sigma$.  The $Z_2$
symmetry $\sigma$ acts by ${\cal I}$ combined with inversion
on the M-theory circle, $x_7 \to -x_7$.
Then for $\mu > 0$, the fixed point loci of $\sigma$,
which consist of copies of $\Sigma^+$ at $x_7 = 0, 1/2$,
are actually $S^3$s of $A_1$ singularities
in $X$.
This gives rise in M-theory on $X$ to two 4d,
${\cal N}=1$ pure $SU(2)$ gauge theories (with equal gauge couplings).

Now, consider taking $\mu$ through 0.  At $\mu \to 0$ there are collapsing
associative three-cycles in $X$, and hence membrane instanton effects are
expected to be large \HM.
However, the sizes of the $S^3$s come paired in chiral multiplets with
periods of the three-form $C$ field over the $S^3$s, and for generic
values of this phase, there is no singularity in the physics --
singularities in ${\cal N}=1$ moduli spaces happen at complex codimension
one.  Therefore, one can smoothly (in the physical sense) continue from
$\mu > 0$ to $\mu < 0$.  This raises a puzzle: the $SU(2)$ gauge groups
present for $\mu > 0$ have now disappeared, since the
$Z_2$ symmetry $\sigma$ acts on $X$ without fixed points for $\mu < 0$.
However, the information about the gauge theory must be encoded somehow
in the $\mu < 0$ geometry.

The basic point is as in \AMV.  For $\mu < 0$, one can still look for
a homologically nontrivial three-sphere which membrane instantons can
wrap.  For instance,  consider the
locus of pure imaginary $z_i$, still at $x_7 = 0,1/2$.
This is given by a three-sphere
\eqn\newsphere{\sum_{i=1}^{4} y_i^2 = - \mu}
which is orbifolded by the freely-acting $Z_2$ symmetry
$y_i \to -y_i$.
Call the resulting $\IR \IP^{3}$ ~$\Sigma^-$.  It turns out that
$\Sigma^{-}(-\mu)$ has exactly half the volume of $\Sigma^+ (\mu)$, due
to the orbifolding.
These $\IR \IP^{3}$s are associative three-cycles in $X$ for $\mu < 0$.
However, as in \AMV, changing the period of the $C$ field on
$\Sigma^{+}(\mu)$ by $2\pi$, which is physically meaningless,
corresponds to changing it by $\pi$ on $\Sigma^{-}(-\mu)$, due to the
smaller volume.
This ambiguity in the choice of phase for $\mu < 0$ corresponds to
the vacuum degeneracy due to the gaugino condensate in the gauge
theory.\foot{Notice that since the two associative $S^3$s at $x_7 = 0,
1/2$ are
in the same homology class, their volumes (and the periods of the
$C$-field)
are the same.  So although there are two $SU(2)$s, the choice of phase
in the two gaugino condensates is related -- there is only a single
$Z_2$ ambiguity.  This carries over to the $IIA$ picture as well.}

In the IIA picture, with $x_7$ taken as the M-theory circle,
this becomes an example where an orientifold plane and two $D6$
branes, present for $\mu > 0$, disappear as $\mu$ passes through 0.
This system has an $SO(4)$ gauge symmetry, and
should give rise to
multiple vacua after gaugino condensation, in agreement with the
M-theory picture above.  The phase ambiguity detected by membrane
instantons in M-theory is detected by $D2$ brane instantons in the
string theory picture.  This is in accord with our gauge theory
intuition, since $D2$ branes are the instantons of the $D6$ brane
gauge theory in the phase where the $D6$ branes exist \douglas.
The fact that $\Sigma^{-}(-\mu)$ has half the volume of
$\Sigma^{+}(\mu)$ then becomes the familiar fact that
the superpotential from a gaugino condensate in ${\cal N}=1$ $SU(2)$
gauge theory looks
like a ``half-instanton effect."

More precisely, once we have compactified this setup, the superpotential
we are discussing destabilizes the closed-string modulus $\mu$ (which
is a parameter in the non-compact case).  In our discussion here,
we are imagining that we can hold $\mu$ fixed at various values,
which is reasonable as long
as the scale generated by the superpotential is parametrically smaller
than the string/Planck scale. This is true for large enough $\vert
\mu \vert$.

It is clear that the other examples of \kklmII, which involve
sL three-cycles $\Sigma$ with $b_1 (\Sigma) > 0$,
could also be lifted in this way to find examples of M-theory
``dualities'' in gauge theories with adjoint matter.
Some examples of this have already appeared in \Cachazo.

\newsec{Discussion}

Little is known about M-theory compactification on spaces of $G_2$
holonomy. Naive extrapolation of the kinds of results that
exist so far suggests that further
study of the relationship between IIA orientifolds and M-theory
compactifications could yield:

\item{1)} A large class of examples of
non-perturbative dualities between orientifolds of
type II compactifications on
Calabi-Yau spaces of different topologies, with
different numbers of space-filling D-branes.

\item{2)} New gauge theory/gravity dualities along the lines of \AMV,
in a compact context (i.e., coupled to 4d gravity).

\item{3)} Connections between the study of disc instanton effects in
type II compactifications with branes
(see e.g. \refs{\kklmI,\kklmII,\ov,\av})
and
membrane instanton effects in M-theory \HM.
The real involution of the CY pairs holomorphic discs \kklmII\ even
when the sL three-cycle on which they
end is deformed away from
the real slice.  In this way, pairs of
discs times the M-theory circle form closed orbifold-invariant
three-manifolds which membrane instantons can wrap.  Similarly,
$\IR \IP^2$ worldsheets with their crosscap on the real slice
lift to orbifolds of membrane instantons on the M-theory circle times the
covering sphere of the $\IR \IP^2$ \SinVa.

\item{4)} A good understanding of the new physics which
arises at singularities of M-theory on spaces of $G_2$ holonomy (some
examples of this were discussed in \Grossfest).  It would
be particularly interesting to find various singularities which
correspond to chiral gauge theories.
Perhaps these would provide a useful tool for the further exploration
of chirality changing
phase transitions \chir.

\item{5)} A new window into type I compactifications.
The type IIA orientifolds studied here are T-dual to type I
string compactifications (roughly speaking, by T-duality on the
$T^3$ fibers \SYZ\ of the
Calabi-Yau space which is being orientifolded).  Therefore,
any insights gained about these
models through their M-theory interpretation will carry over to
the study of certain type I theories.

\bigskip
\centerline{\bf{Acknowledgements}}
\medskip

We would like to thank J. Polchinski and
E. Silverstein for helpful discussions, and B. Acharya for pointing
out a minor error in \S5\ of the first version of this note.
This work was supported in part by NSF grant PHY-95-14797 and
by the DOE under contract DOE-AC03-76SF00098.
S.K. enjoyed the hospitality of the ITP Santa Barbara
``M-theory'' program while performing the work reported here,
and was supported by the National Science Foundation under
grant number PHY-99-07949.
The work of S.K. was supported in part by a David and Lucile Packard
Foundation Fellowship for Science and Engineering and an Alfred P.
Sloan Foundation Fellowship.
The work of J. M. was supported in part by the Department of Defense
NDSEG Fellowship program.

\listrefs
\end